\documentclass[times, 10pt,twocolumn]{article} 
\usepackage{times}
\usepackage{latex8}
\usepackage[american]{babel}
\usepackage{amsmath}
\usepackage{amssymb}
\usepackage{graphicx}
\usepackage{color}
\usepackage[boxruled]{algorithm2e}

\definecolor{lightgray}{gray}{0.95}
\definecolor{darkgray}{gray}{0.65}

\newcommand{\equals}{\stackrel{\mathrm{def}}{=}}

\newcommand{\dm}{\mbox{\textsf{DM}}} 
\newcommand{\edf}{\mbox{\textsf{EDF}}}
\newcommand{\agr}{\mbox{\textsf{AGR}}}
\newcommand{\mora}{\mbox{\textsf{MORA}}}
\newcommand{\mote}{\mbox{\textsf{MOTE}}}
\newcommand{\maxmethod}{\mbox{\textsf{MAX}}}
\newcommand{\off}{\mbox{\textsf{OFF}}}
\newcommand{\moraote}{\mbox{\textsf{MORAOTE}}}
\newcommand{\dra}{\mbox{\textsf{DRA}}}

\newcommand{\somme}{\operatorname{sum}}
\newcommand{\offline}{\operatorname{off}}
\newcommand{\rem}{\operatorname{rem}}
\newcommand{\next}{\operatorname{next}}
\newcommand{\disp}{\operatorname{disp}}
\newcommand{\idle}{\operatorname{idle}}
\newcommand{\nextdisp}{\operatorname{nextdisp}}

\newcommand{\act}{\operatorname{actual}}

\newcommand{\arem}{\rem^{\offline}}
\newtheorem{Theorem}{Theorem}

\newtheorem{Lemma}{Lemma}

\newtheorem{Definition}{Definition}
\newtheorem{ARule}{$\alpha$-Rule}
\newtheorem{Rule}{Rule}

\newtheorem{Observation}{Observation}

\newenvironment{proof}[1][Proof]{\begin{trivlist}
\item[\hskip \labelsep {\bfseries #1}]}{\end{trivlist}}

\newcommand{\qed}{\rule{7pt}{7pt}}

\newcommand{\deadline}[1]{
\begin{picture}(1,3)
\put(0,1){\circle{0.5}}
\put(0,0){\makebox(1,1)[l]{#1}}
\end{picture}
}

\title{$\mora$: an Energy-Aware Slack Reclamation Scheme for \\ Scheduling Sporadic Real-Time Tasks upon Multiprocessor Platforms}
\author{Vincent Nelis$^{1,2}$ \and Jo\a"el Goossens$^1$}
\begin{document}
\maketitle
\footnotetext[1]{Universit\'e Libre de Bruxelles (U.L.B.) CP 212, Department of Computer Science, 50 Av. F.D.~ Roosevelt, B-1050 Brussels, Belgium.}
\addtocounter{footnote}{1}
\footnotetext[2]{Supported by the Belgian National Science Foundation (F.N.R.S.) under a F.R.I.A. grant.}
\addtocounter{footnote}{1}

\begin{abstract}
In this paper, we address the global and preemptive energy-aware scheduling problem of sporadic constrained-deadline tasks on DVFS-identical multiprocessor platforms. We propose an online slack reclamation scheme which profits from the discrepancy between the worst- and actual-case execution time of the tasks by slowing down the speed of the processors in order to save energy. Our algorithm called $\mora$ takes into account the application-specific consumption profile of the tasks. We demonstrate that $\mora$ does not jeopardize the system schedulability and we show by performing simulations that it can save up to $32 \%$ of energy (in average) compared to execution without using any energy-aware algorithm. 
\end{abstract}

\Section{Introduction}

\textbf{Context of the study.} Nowadays, many modern processors can operate at various supply voltages, where different supply voltages lead to different clock frequencies and to different processing speeds. Since the power consumption of a processor is usually a convex and increasing function of its speed, the slower its speed is, the less its consumption is~\cite{Kuo:06}. Among the most recent and popular such processors, one can cite the Intel PXA27x processor family~\cite{IntelDatasheet:05}, used by many PDA devices~\cite{UrlDevice}. 

Many computer systems, especially embedded systems, are now equipped with such voltage (speed) scaling processors and adopt various energy-efficient strategies for managing their applications intelligently. Moreover, many recent energy-constrained embedded systems are built upon multiprocessor platforms because of their high-computational requirements. As pointed out in~\cite{BaAn:03,AndersonBaruah:04}, another advantage is that multiprocessor systems are more energy efficient than equally powerful uniprocessor platforms, because raising the frequency of a single processor results in a \emph{multiplicative} increase of the consumption while adding processors leads to an \emph{additive} increase. 

Supported by this emerging technology, the Dynamic Voltage and Frequency Scaling (DVFS)~\cite{Chen:07} framework becomes a major concern for multiprocessor power-aware embedded systems. For real-time systems, this framework consists in reducing the system energy consumption by adjusting the working voltage and frequency of the processors, while respecting all the timing constraints.

\textbf{Previous work.} There are a large number of researches about the \emph{uni}processor energy-aware real-time scheduling problem~\cite{AyMeMoMe:04,Bansal:04,Irani:03,IsYa:98,YaDeSh:95}. Among those, many \emph{slack reclamation} approaches have been developed over the years. Such techniques dynamically collect the unused computation times at the end of each early task completion and share it among the remaining pending tasks. Examples of such approaches include the ones proposed in~\cite{AyMeMoMe:04,pillai01,ShCh:99,ZhCh:02}. Some reclaiming algorithms even anticipate the early completion of tasks for further reducing the CPU speed~\cite{AyMeMoMe:04,pillai01}, some having different levels of ``aggressiveness''~\cite{AyMeMoMe:04}.

In~\cite{Chen:07}, Kuo et al.\@ propose a state-of-art about energy-aware algorithms in \emph{multi}processor environment. As it is mentioned in this state-of-art, many studies (see for instance~\cite{ChHsChYaPaKu:04,ChenKuo:05,Kuo:06,ChenKuo:06,ZhuMelhemChilders:01,YangChenKuo:05}) consider the \emph{frame-based task model}, i.e., all the tasks share a common deadline and this ``frame'' is indefinitely repeated. Among the most interesting studies which consider this task model, Zhu et al.\@~\cite{ZhuMelhemChilders:01} explored \emph{online} slack reclamation schemes (i.e., running during the system execution) for dependent and independent tasks. In~\cite{ChenKuo:06}, Kuo et al.\@ propose a set of energy-efficient scheduling algorithms with different task remapping and slack reclamation schemes. In~\cite{Kuo:06}, the authors address independent tasks, where task migrations are not allowed. In~\cite{ChHsChYaPaKu:04}, the authors provide some techniques with and without allowing task migration, while assuming that tasks share the same power consumption function and each processor may run at a selected speed, independently from the speeds of the others. In~\cite{ChenKuo:05}, the authors consider that tasks are allowed to have different power consumption functions. In~\cite{YangChenKuo:05}, energy-aware multiprocessor scheduling of frame-based tasks was explored for multiprocessor architectures, in which all the processors must share the same speed at any time. Finally the authors of~\cite{BertenGoossens:08} propose a slack reclamation scheme for identical multiprocessor platforms, while considering frame-based tasks of which the distribution of the computation times is assumed to be known.

Targeting a \emph{sporadic task model}, Anderson and Baruah~\cite{AndersonBaruah:08} explored the trade-off between the total energy consumption of task executions and the number of required processors, where all the tasks run at the same common speed. In previous work~\cite{Nelis:08}, we provided a technique that determines the minimum common offline speed for every task under global-\edf\ policy~\cite{Baker:03}, while considering identical multiprocessor platforms. Furthermore, we proposed in the same study an \emph{online} algorithm called $\mote$ which was, to the best of our knowledge, the first to address the global and preemptive energy-aware scheduling problem of sporadic constrained-deadlines tasks on multiprocessors. The main idea of $\mote$ is to anticipate at run-time the coming idle instants in the schedule in order to reduce the processors speed accordingly. This algorithm cannot be considered as a slack reclamation scheme since it does not directly take advantage from early tasks completion, but it can be combined with slack reclaiming techniques (and in particular with $\mora$) in order to improve the energy savings.

\textbf{Contribution of the paper.} In this paper, we propose a \emph{slack reclamation scheme} called $\mora$ for the global and preemptive energy-aware scheduling problem of \emph{sporadic constrained-deadline real-time tasks} on a \emph{fixed number of DVFS-capable processors}. According to~\cite{Chen:07} and to the best of our knowledge, this is the first work which addresses a slack reclamation scheme in this context. Although most previous studies on multiprocessor energy-efficient scheduling assumed that the actual execution time of a task is equal to its Worst-Case Execution Time (WCET), such that those in~\cite{AlEnawyAydin:05,AydinYang:03,ChHsChYaPaKu:04,YangChenKuo:05} for instance, this work is motivated by the scheduling of tasks in practice, where tasks might usually complete earlier than their WCET~\cite{AyMeMoMe:04,ZhuMelhemChilders:01}. The proposed algorithm $\mora$ is an \emph{online} scheme which exploits early task completions by using as much as possible the unused time to reduce the speed of the processors. Although it has been inspired from the \emph{uni}processor ``Dynamic Reclaiming Algorithm'' ($\dra$) proposed in~\cite{AyMeMoMe:04}, the way in which it profits from the unused time is very different from the $\dra$ since $\mora$ takes into account the application-specific consumption profile of the tasks.

\textbf{Organization of the paper.} The document is organized as follows: in
Section~\ref{sec:model}, we introduce our model of computation, in particular our task and platform model; 
in Section~\ref{sec:mora}, we present our online slack reclamation technique called $\mora$ and we prove its correctness; 
in Section~\ref{sec:experiments}, we present our simulation results
and in Section~\ref{sec:conclusion}, we introduce future research directions and we conclude. 

\Section{Model of computation}
\label{sec:model}

\SubSection{Platform model}
\label{sec:platform}

\begin{table}
\centering
\begin{footnotesize}
\begin{tabular}{| r || c | c | c | c | c |}
\hline
Processor Type & \multicolumn{5}{ c |}{Intel XScale~\cite{IntelXScale}} \\
\hline
Frequency (MHz): $f_k$ & 150 & 400 & 600 & 800 & 1000 \\
\hline 
Speed: $s_k$ & 0.15 & 0.4 & 0.6 & 0.8 & 1.0 \\
\hline
Voltage (V) & 0.75 & 1.0 & 1.3 & 1.6 & 1.8 \\
\hline
Power in run mode (mW): $P(s_k)$ & 80 & 170 & 400 & 900 & 1600 \\
\hline
Power in idle mode (mW): $P_{\idle}$ & \multicolumn{5}{ c |}{40} \\
\hline
\end{tabular}
\end{footnotesize}
\label{tab:processor_characterisitc}
\caption{Intel XScale characteristics}
\vspace{-1ex}
\end{table}

We consider multiprocessor platforms composed of a known and fixed number $m$ of DVFS-\emph{identical} processors $\left\{ {\cal P}_1, {\cal P}_2, \ldots, {\cal P}_m \right\}$. ``DVFS-identical'' means that (i) all the processors have the same profile (in term of consumption, computational capabilities, etc.) and are interchangeable, (ii) two processors running at a same frequency execute the same amount of execution units, and (iii) all the processors have the same minimal and maximal operating frequency denoted by $f_{\min}$ and $f_{\max}$, respectively. The processors are referred to as \emph{independent}, with the interpretation that they can operate at different frequencies at the same time~\cite{Talpes:05,Magklis:03}. Furthermore, we assume that each processor can dynamically adapt its operating frequency (and voltage) at any time during the system execution, independently from each other. The time overheads on frequency (voltage) switching are assumed to be negligible, such as in many researches~\cite{AydinMelhem:01,Bansal:04,Levner:04,YaDeSh:95,Zhuo:05}.

We define the notion of \emph{speed} $s$ of a processor as the ratio of its operating frequency $f$ over its maximal frequency, i.e.: $s \equals \frac{f}{f_{\max}}$ -- with the interpretation that a job that executes on a processor running at speed $s$ for $R$ time units completes $s \times R$ execution units. When only $K$ discrete frequencies are available to a processor, they are sorted in the increasing order of frequency and denoted by $f_1, \ldots, f_K$. For each frequency $f_k$ such that $1 \leq k \leq K$, we denote by $s_k$ the corresponding speed (i.e., $s_k \equals \frac{f_k}{f_{\max}}$) and by $P(s_k)$ the power consumption (energy consumption rate) per second while the processor is running at speed $s_k$. The available frequencies and the corresponding core voltages of the Intel XScale processor~\cite{IntelXScale} that will be used in our experiments are outlined in Table 1. Notice that, from our definition of the processor speed, $s_{\max}$ is $\frac{f_{\max}}{f_{\max}} = 1$ whatever the considered processor. Moreover, due to the finite number of speeds that are available to any practical processor, any speed $s$ computed by any energy-aware algorithm must be translated into one of the available speeds. In this work, this translation is performed by the function $\widehat{S}(s) \equals \min\{s_i \mid s_i \geq s\}$.


\SubSection{Application model}

A real-time system $\tau$ is a set of $n$ functionalities denoted by $\left\{ \tau_1, \tau_2, \ldots, \tau_{n} \right\}$. Every functionality $\tau_i$ is modeled by a \emph{sporadic constrained-deadline} task characterized by three parameters $(C_i, D_i, T_i)$ -- a Worst-Case Execution Time (WCET) $C_{i}$ {at maximal processors speed $s_{\max}$} (expressed in milliseconds for instance), a minimal inter-arrival delay $T_{i}$ and a relative deadline $D_{i} \leq T_i$ -- with the interpretation that the task $\tau_i$ generates successive \emph{jobs} $\tau_{i,j}$ (with $j = 1, \ldots, \infty$) arriving at times $a_{i,j}$ such that $a_{i,j} \geq a_{i,j-1} + T_i$ (with $a_{i,1} \geq 0$), each such job has a worst-case execution time of at most $C_{i}$ time units (at maximal processors speed $s_{\max}$), and must be completed at (or before) its absolute deadline noted $D_{i,j} \equals a_{i,j} + D_i$. According to our definition of the processors speed, a processor running at speed $s_{\max} = 1$ may take up to $C_i$ time units to complete a job $\tau_{i,j}$ and, at a given speed $s$, its WCET is $\frac{C_i}{s}$. Notice that, since $D_i \leq T_i$, successive jobs of any task $\tau_i$ do not interfere with each other. 

We define the \emph{density} $\delta_i$ of the task $\tau_i$ as the ratio of its WCET at maximal speed $s_{\max}$ over its deadline, i.e., $\delta_i \equals \frac{C_i}{D_i}$. We assume that this ratio is not larger than $1$ for every task, since a task with a density larger than $1$ is never able to meet its deadlines (since task parallelism is forbidden in this work). The \emph{maximal density} $\delta_{\max}(\tau)$ \emph{of the system} is defined as $\delta_{\max}(\tau) \equals \max_{i=1}^{n}\{ \delta_i \}$ and its \emph{total density} is defined as $\delta_{\somme}(\tau) \equals \sum_{i=1}^n \delta_{i}$. In our study, all the tasks are assumed to be \emph{independent}, i.e., there is no communication, no precedence constraint and no shared resource (except the processors) between them. 

At any time $t$ in any schedule ${\cal S}$, a job $\tau_{i,j}$ is said to be \emph{active} iff $a_{i,j} \leq t$ and it is not completed yet in ${\cal S}$. Moreover, an active job is said to be \emph{running} at time $t$ in ${\cal S}$ if it is executing on a processor. Otherwise, the active job is pending in a ready-queue of the operating system and we say that it is \emph{waiting}. Furthermore, a job is said to be \emph{dispatched} at time $t$ in ${\cal S}$ if it passes from the waiting state to the running state at time $t$.

Although certain benchmarks provide measured power consumption, we should not ignore that different applications may have different instruction sequences and require different function units in the processor, thus leading to different dynamic consumption profiles. As it was already done in~\cite{RXu:07}, we hence introduce a measurable parameter $e_i$ for each task $\tau_i$ that reflects this application-specific power difference between the applications and the measured benchmark. Accordingly, the consumption of any task $\tau_i$ executed for $1$ time unit at speed $s_k$ can be estimated by $e_i \cdot (P(s_k) - P_{\idle})+P_{\idle}$~\cite{RXu:07}, where $P(s)$ and $P_{\idle}$ are defined as in Table 1. In the remainder of this paper, we denote by $E_i(R, s_k)$ the energy consumed by the task $\tau_i$ when executed for $R$ time units at speed $s_k$ and we define it as $E_i(R, s_k) \equals R \cdot (e_i \cdot (P(s_k) - P_{\idle})+P_{\idle})$. As we will see in Section~\ref{sec:principemora}, $\mora$ uses these energy consumption functions in order to improve the energy saving that it provides. This improvement makes $\mora$ very different from the \emph{uni}processor dynamic reclaiming algorithm $\dra$ proposed in~\cite{AyMeMoMe:04}.

\SubSection{Scheduling specifications}
\label{sec:scheduler}

We consider in this study the \emph{global} scheduling problem of sporadic constrained-deadlines tasks on multiprocessor platforms. ``Global'' scheduling algorithms, on the contrary to partitioned algorithms, allow different tasks and \emph{different} jobs of the same task to be executed upon \emph{different} processors. Furthermore, we consider \emph{preemptive} scheduling and \emph{Fixed Job-level} Priority assignment (FJP), with the following interpretations. In the \emph{preemptive global scheduling problem}, every job can start its execution on any processor and \emph{may migrate at run-time} to any other processor if it gets meanwhile preempted by a higher-priority job. We assume in this paper that preemptions are carried out with no loss or penalty. \emph{Fixed Job-level Priority assignment} means that the scheduler assigns a priority to jobs as soon as they arrive and every job keeps its priority constant until it completes. \emph{Global Deadline Monotonic} and \emph{Global Earliest Deadline First}~\cite{Baker:03} are just some examples of such scheduling algorithms. 

\Section{The Multiprocessor Online Reclaiming Algorithm (\mora)}
\label{sec:mora} 

\SubSection{Notations}
\label{sec:notations}
\emph{During the system execution}, every \emph{active} job $\tau_{i,j}$ has two associated speeds noted $s_{i,j}$ and $s_{i,j}^{\offline}$. The speed $s_{i,j}$ denotes \emph{the speed that a processor adopts while executing $\tau_{i,j}$}. We assume that these execution speeds $s_{i,j}$ can be modified at any time during the system execution, even during the execution of $\tau_{i,j}$, and it is instantaneously reflected on the processor speed. On the other hand, the speed $s_{i,j}^{\offline}$ is the \emph{offline precomputed execution speed} of $\tau_{i,j}$, in the sense that the value of $s_{i,j}$ is always set to $s_{i,j}^{\offline}$ at $\tau_{i,j}$ arrival time. These offline speeds $s_{i,j}^{\offline}$ are determined \emph{before} the system execution and remain always constant at run-time. They may be simply set to the maximal processors speed $s_{\max}$, or they can be determined by an \emph{offline} energy-aware strategy, such that the one proposed in~\cite{Nelis:08} for instance. These offline speeds \emph{must ensure that all the deadlines are met} when the set of tasks is scheduled upon the $m$ processors, even if every job of every task presents its WCET. Notice that, since each task generates an infinity of jobs, the method proposed in~\cite{Nelis:08} determines a common speed for every task and assumes that every job $\tau_{i,j}$ inherits from the offline speed of $\tau_i$ at run-time.

$\mora$ is based on reducing \emph{online} (i.e., while the system is running) the execution speed $s_{i,j}$ of the jobs in order to provide energy savings \emph{while still meeting all the deadlines}. To achieve this goal, $\mora$ detects whenever the speed $s_{i,j}$ of an active job $\tau_{i,j}$ can safely be reduced by performing comparison between the schedule which is actually produced (called the \emph{actual schedule} hereafter) and the \emph{offline schedule} defined below. We will see in the remainder of this section that our algorithm $\mora$ \emph{always refers to this offline schedule} in order to produce the actual one.

\begin{Definition}[The offline schedule]
The offline schedule is the schedule produced by the considered scheduling algorithm on which every job of every task $\tau_i$ runs at its offline speed $s_{i,j}^{\offline}$ and presents its WCET.
\end{Definition}

\begin{figure}[h!]
\centering
  {\footnotesize \setlength{\unitlength}{0.37cm}
\begin{picture}(22,14)
\put(1,0){\vector(1,0){21}}
\put(21,-1){time}
\put(1.9,-0.5){\tiny $0$}
\put(2.9,-0.5){\tiny $1$}
\put(3.9,-0.5){\tiny $2$}
\put(4.9,-0.5){\tiny $3$}
\put(5.9,-0.5){\tiny $4$}
\put(6.9,-0.5){\tiny $5$}
\put(7.9,-0.5){\tiny $6$}
\put(8.9,-0.5){\tiny $7$}
\put(9.9,-0.5){\tiny $8$}
\put(10.9,-0.5){\tiny $9$}
\put(11.9,-0.5){\tiny $10$}
\put(12.9,-0.5){\tiny $11$}
\put(13.9,-0.5){\tiny $12$}
\put(14.9,-0.5){\tiny $13$}
\put(15.9,-0.5){\tiny $14$}
\put(16.9,-0.5){\tiny $15$}
\put(17.9,-0.5){\tiny $16$}
\put(18.9,-0.5){\tiny $17$}
\put(19.9,-0.5){\tiny $18$}

\put(0,13){(a) Offline schedule.}
\put(0,10.5){${\cal P}_1$}
\put(0,9){${\cal P}_2$}
\multiput(2,8)(1,0){21}{
     \multiput(0,0)(0,0.35){11}{\tiny $\cdot$}
}
\put(5.1,10){$\overbrace{\makebox(3,1){}}^{\arem_{1,1}(3)}$}
\put(2.1,10){\framebox(6,1)[c]{$\tau_{1,1}$}}
\put(8.1,10){\framebox(8,1)[c]{$\tau_{3,1}$}}
\put(8.1,11.6){\line(0,-1){.7}}
\put(8.4,11.3){\tiny $\disp_{3,1}(0)$}
\put(2.1,8.6){\framebox(6,1)[c]{$\tau_{2,1}$}}
\put(8.1,8.6){\framebox(2,1)[c]{$\tau_{4,1}$}}
\put(8.1,7.9){\line(0,1){.7}}
\put(5.5,7.3){\tiny $\disp_{4,1}(0)$}
\put(10.1,8.6){\framebox(6,1)[c]{$\tau_{5,1}$}}
\put(10.1,7.9){\line(0,1){.7}}
\put(9.5,7.3){\tiny $\disp_{5,1}(0)$}

\put(20.1,7.7){\deadline}
\put(17.1,7.7){deadline of $\tau_{5,1}$}

\put(0,6){(b) Actual schedule.}
\put(0,4){${\cal P}_1$}
\put(0,2.5){${\cal P}_2$}
\multiput(2,1)(1,0){21}{
     \multiput(0,0)(0,0.35){13}{\tiny $\cdot$}
}
\put(2.1,3.5){\framebox(3,1)[c]{$\tau_{1,1}$}}
\put(8.1,3.5){\framebox(3,1)[c]{$\tau_{3,1}$}}
\put(2.1,2.1){\framebox(2,1)[c]{$\tau_{2,1}$}}
\put(8.1,2.1){\framebox(2,1)[c]{$\tau_{4,1}$}}
\put(10.1,2.1){\framebox(6,1)[c]{$\tau_{5,1}$}}
\put(20.1,1.2){\deadline}
\put(17.1,1.2){deadline of $\tau_{5,1}$}
\end{picture}
}
\caption{Offline and actual schedules.}
\label{fig:notations}
\end{figure}
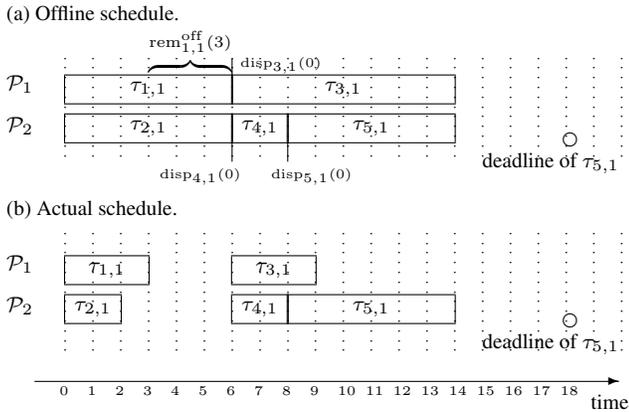

Figure~\ref{fig:notations}.(a) depicts an example of an offline schedule and illustrates the notations that will be used throughout the paper. In this picture, a 5-tasks system is executed upon 2 processors, where only the first job of each task is represented. The characteristics of the tasks are the following (remember that $\tau_i = (C_i, D_i, T_i)$): $\tau_1 = (6, 14, 30), \tau_2 = (6, 15, 35), \tau_3 = (8, 16, 40), \tau_4 = (2, 17, 45)$ and $\tau_5 = (6, 18, 50)$. Assuming Global-$\edf$, we have the following priority order: $\tau_{1,1} > \tau_{2,1} > \tau_{3,1} > \tau_{4,1} > \tau_{5,1}$. Furthermore, we assume in this example that the offline speed $s_{i,j}^{\offline}$ of every job $\tau_{i,j}$ is the maximal processors speed $s_{\max} = 1$.

At run-time, \emph{whenever any job is dispatched to any processor ${\cal P}_{\ell}$ in the offline schedule, $\mora$ also dispatches it to ${\cal P}_{\ell}$ in the actual one}. That is, assuming the same set of tasks as in Figure~\ref{fig:notations}.(a), Figure~\ref{fig:notations}.(b) depicts the actual schedule that is produced if the \emph{actual execution time} of the jobs $\tau_{1,1}, \ldots, \tau_{5,1}$ are respectively $3, 2, 3, 2, 6$. At any time $t$, we denote by $\rem_{i,j}(t)$ and $\arem_{i,j}(t)$ the worst-case \emph{remaining} execution time of job $\tau_{i,j}$ at speed $s_{\max}$ in the actual and offline schedule, respectively. We assume that these quantities are updated at run-time for every active job $\tau_{i,j}$. For instance in Figure~\ref{fig:notations} at time $t=3$, we have $\rem_{1,1}(3) = 0$ (since $\tau_{1,1}$ completes at time $t=3$ in the actual schedule) and $\arem_{1,1}(3) = 3$. Notice that from our definition of a processor speed, the worst-case \emph{remaining} execution time of job $\tau_{i,j}$ at speed $s$ in the actual and offline schedule is $\frac{\rem_{i,j}(t)}{s}$ and $\frac{\arem_{i,j}(t)}{s}$, respectively. We denote by $\disp_{i,j}(t)$ the earliest time at which $\tau_{i,j}$ is dispatched in the offline schedule, when only the set of active jobs at time $t$ in the offline schedule are considered. For instance in Figure~\ref{fig:notations}.(a) we have $\disp_{4,1}(0) = 6$ and $\disp_{5,1}(0) = 8$. Finally, $\nextdisp({\cal P}_{\ell}, t)$ denotes the earliest instant after time $t$ at which a job which is not completed in the actual schedule at time $t$ is dispatched to ${\cal P}_{\ell}$ in the offline schedule. Again, only the set of active jobs at time $t$ in the offline schedule are considered to compute $\nextdisp({\cal P}_{\ell}, t)$. For instance in Figure~\ref{fig:notations}.(a) we have $\nextdisp({\cal P}_2, 2) = 6$ and $\nextdisp({\cal P}_1, 3) = 6$.

\SubSection{The $\alpha$-queue}

Since the jobs arrival times are unknown while considering the sporadic task model, computing and storing the entire offline schedule cannot be done \emph{before the system execution}. Hence, our algorithm only stores and updates at run-time a \emph{sufficient part} of the offline schedule. This kind of approach (i.e., using a dynamic data structure for embodying a sufficient part of the offline schedule) was previously proposed in~\cite{AyMeMoMe:04}. As in~\cite{AyMeMoMe:04}, we call this data structure \emph{$\alpha$-queue}. The \emph{$\alpha$-queue} is a list that contains, at any time $t$, the worst-case remaining execution time $\arem_{i,j}(t)$ of every active jobs $\tau_{i,j}$ \emph{in the offline schedule}. This list is managed according to the following rules, which are widely inspired from~\cite{AyMeMoMe:04}.

\begin{ARule}
\label{arule:1}
At any time, the $\alpha$-queue is sorted by decreasing order of the job priorities, with the $m$ highest priority jobs at the head of the queue.
\end{ARule}

\begin{ARule}
\label{arule:2}
Initially the $\alpha$-queue is empty. 
\end{ARule}

\begin{ARule}
\label{arule:3}
Upon arrival of a job $\tau_{i,j}$ at time $t$, $\tau_{i,j}$ inserts its WCET $C_i$ into the $\alpha$-queue in the correct priority position. This happens only once for each arrival, no re-insertion at return from preemptions. 
\end{ARule}

\begin{ARule}
\label{arule:4}
As time elapses, the $m$ fields $\arem_{i,j}(t)$ (if any) at the head of the $\alpha$-queue are decreased with a rate proportional to the offline speeds $s_{i,j}^{\offline}$. Whenever one field reaches zero, that element is removed and the update continues, still with the $m$ first elements (if any). Obviously, no update is performed when the $\alpha$-queue is empty.
\end{ARule}

For the same reasons than those explained in~\cite{AyMeMoMe:04}, the following observation holds.
\begin{Observation}
\label{obs:1}
At any time $t$, the $\alpha$-queue updated according to $\alpha$-Rules~\ref{arule:2}--\ref{arule:4} contains only the jobs that would be active at time $t$ in the offline schedule. Moreover, the $\arem_{i,j}(t)$ fields contain the worst-case remaining execution time of every active job $\tau_{i,j}$ at time $t$ in the offline schedule.
\end{Observation}

By consulting the $\alpha$-queue at any time $t$, $\mora$ is able to get the required information about any active jobs $\tau_{i,j}$ in the offline schedule, i.e., its worst-case remaining execution time $\arem_{i,j}(t)$, its next dispatching time $\disp_{i,j}(t)$ and the next job dispatching time $\nextdisp({\cal P}_{\ell},t)$ on any processor ${\cal P}_{\ell}$. Due to the space limitation, we omitted the implementation details about the procedures which compute $\disp_{i,j}(t)$ and $\nextdisp({\cal P}_{\ell}, t)$.

Notice that, as explained in~\cite{AyMeMoMe:04}, the dynamic reduction of $\arem_{i,j}(t)$ from $\alpha$-Rule~\ref{arule:4} does not need to be performed at every clock cycle. Instead, for efficiency, we perform the reduction only \emph{before} $\mora$ modifies a speed, by taking into account the time elapsed since the last update. Formally, if $\Delta t$ time units elapsed, the $m$ fields at the head of the $\alpha$-queue are updated as follows: $\arem_{i,j}(t + \Delta t) \leftarrow \arem_{i,j}(t) - s_{i,j}^{\offline} \cdot \Delta t$. The above approach relies on two facts: as we will see in the next section, the speed adjustment decisions will be taken only at job arrival time (i.e., the execution speed of the arriving job is set to its offline speed), job dispatching time in the offline schedule and whenever a processor is about to get idle in the actual schedule. Hence, it is necessary to have an accurate $\alpha$-queue only at these instants. Second, between these instants, each task is effectively executed non-preemptively in the actual schedule. 

\begin{figure}[h!]
  {\footnotesize \setlength{\unitlength}{0.37cm}
\begin{picture}(22,11)
\put(1,0){\vector(1,0){21}}
\put(21,-1){time}
\put(1.9,-0.5){\tiny $0$}
\put(2.9,-0.5){\tiny $1$}
\put(3.9,-0.5){\tiny $2$}
\put(4.9,-0.5){\tiny $3$}
\put(5.9,-0.5){\tiny $4$}
\put(6.9,-0.5){\tiny $5$}
\put(7.9,-0.5){\tiny $6$}
\put(8.9,-0.5){\tiny $7$}
\put(9.9,-0.5){\tiny $8$}
\put(10.9,-0.5){\tiny $9$}
\put(11.9,-0.5){\tiny $10$}
\put(12.9,-0.5){\tiny $11$}
\put(13.9,-0.5){\tiny $12$}
\put(14.9,-0.5){\tiny $13$}
\put(15.9,-0.5){\tiny $14$}
\put(16.9,-0.5){\tiny $15$}
\put(17.9,-0.5){\tiny $16$}
\put(18.9,-0.5){\tiny $17$}
\put(19.9,-0.5){\tiny $18$}

\put(0,10.5){(a) Offline schedule.}
\put(0,8.5){${\cal P}_1$}
\put(0,7){${\cal P}_2$}
\multiput(2,6)(1,0){21}{
     \multiput(0,0)(0,0.35){11}{\tiny $\cdot$}
}

\put(2.1,8){\framebox(6,1)[c]{$\tau_{1,1}$}}
\put(8.1,8){\framebox(8,1)[c]{$\tau_{3,1}$}}
\put(2.1,6.6){\framebox(6,1)[c]{$\tau_{2,1}$}}
\put(8.1,6.6){\framebox(2,1)[c]{$\tau_{4,1}$}}
\put(10.1,6.6){\framebox(6,1)[c]{$\tau_{5,1}$}}
\put(20.1,5.7){\deadline}
\put(17.1,5.7){deadline of $\tau_{5,1}$}

\put(0,5){(b) Actual schedule.}
\put(0,3){${\cal P}_1$}
\put(0,1.5){${\cal P}_2$}
\multiput(2,0)(1,0){21}{
     \multiput(0,0)(0,0.35){13}{\tiny $\cdot$}
}

\put(2.1,3){\framebox(3,1)[c]{$\tau_{1,1}$}}
\put(8.1,3){\framebox(3,1)[c]{$\tau_{3,1}$}}
\put(2.1,1.6){\framebox(2,1)[c]{$\tau_{2,1}$}}
\put(4.1,1.6){\framebox(4,1)[c]{$\tau_{5,1}$}}
\put(4.1,1.6){$\underbrace{\makebox(4,1){}}_{L_{4,1}(2)}$}
\put(8.1,1.6){\framebox(2,1)[c]{$\tau_{4,1}$}}
\put(10.1,1.6){\framebox(6,1)[c]{$\tau_{5,1}$}}
\put(20.1,0.7){\deadline}
\put(17.1,0.7){deadline of $\tau_{5,1}$}

\end{picture}
}
\caption{Rules~\ref{rule:1} and~\ref{rule:2} of $\mora$.}
\label{fig:first_rule}
\end{figure}
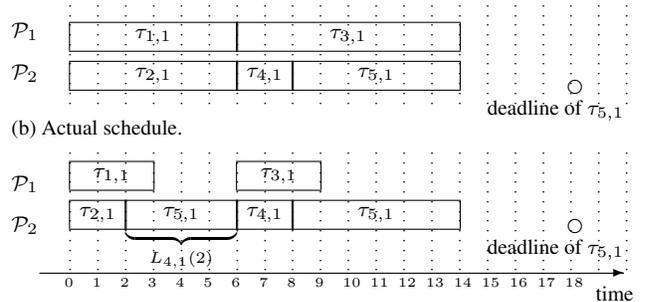

\SubSection{Principle of $\mora$}
\label{sec:principemora}
As explained in Section~\ref{sec:notations}, whenever a job is dispatched in the offline schedule, it is also dispatched in the actual one. However, as we will see below, $\mora$ profits from an early job completion by starting the execution of some other jobs \emph{earlier in the actual schedule than in the offline one}. As a result,  when a job (say $\tau_{k,\ell}$) is dispatched at time $t$ in the offline schedule (and thus also in the actual one), its worst-case remaining execution time $\rem_{k,\ell}(t)$ could be lower than $\arem_{k,\ell}(t)$ if it was executed earlier in the actual schedule. For example, Figure~\ref{fig:first_rule} depicts the same set of tasks than in Figure~\ref{fig:notations}. At time $t=2$, $\tau_{2,1}$ completes in the actual schedule on processor ${\cal P}_2$ and leaves $4$ unused time units. These $4$ time units are reclaimed by starting the execution of $\tau_{5,1}$ (we will see below how $\mora$ selects the job which profits from the slack time) and therefore, when $\tau_{5,1}$ is dispatched to ${\cal P}_2$ in the offline schedule at time $t=8$, it is also dispatched to ${\cal P}_2$ in the actual one and we have $\rem_{5,1}(8) < \arem_{5,1}(8)$. The difference between these remaining execution times is called the \emph{earliness} of the job and we denote it by $\epsilon_{k,\ell}(t) \equals \arem_{k,\ell}(t) - \rem_{k,\ell}(t)$. According to this earliness, whenever any job $\tau_{k,\ell}$ is dispatched in both schedules, its execution speed $s_{k,\ell}$ may \emph{safely} be reduced to $s_{k,\ell}'$ so that $\frac{\rem_{k,\ell}}{s_{k,\ell}'} = \frac{\rem_{k,\ell} + \epsilon_{i,j}(t)}{s_{k,\ell}^{\offline}}$. Indeed, under this speed $s_{k,\ell}'$, $\tau_{k,\ell}$ would complete simultaneously in both schedules if it presents its WCET. This leads to the first rule of $\mora$.

\begin{small}
\begin{Rule}
\label{rule:1}
Any job $\tau_{i,j}$ which is dispatched to any processor ${\cal P}_{\ell}$ at time $t$ in the offline schedule is also dispatched to ${\cal P}_{\ell}$ at time $t$ in the actual one and its execution speed $s_{i,j}$ is modified according to
\begin{equation}
\label{equ:speedreduction2}
s_{i,j} \leftarrow \widehat{S}\left(\frac{\rem_{i,j}(t) \cdot s_{i,j}^{\offline}}{\arem_{i,j}(t)} \right)
\end{equation}
\end{Rule}
\end{small}

The main idea of $\mora$ can be summarized as follows. \emph{When any job completes in the actual schedule without consuming its WCET, the unused time may be reclaimed by starting the execution of any waiting job earlier; and since this waiting job receives additional time for its execution, it can thereby reduce its execution speed}. Using this concept, Figure~\ref{fig:first_rule} depicts an example of how $\mora$ takes advantage from an early job completion. When $\tau_{2,1}$ completes at time $t=2$ in the actual schedule, $\mora$ selects a waiting job (here, $\tau_{5,1}$) and executes it during the $4$ time units left by $\tau_{2,1}$. Since $\tau_{5,1}$ is granted to use $4$ additional time units, $\mora$ reduces its execution speed $s_{5,1}$ so that its worst-case remaining execution time increases by $4$ time units. The selected job is the one for which the resulting speed reduction leads to the highest energy saving. Formally, $\mora$ selects a waiting job and decreases its execution speed as described by Rule~\ref{rule:2}.

\begin{small}
\begin{Rule}
\label{rule:2}
Whenever any processor ${\cal P}_r$ is about to get idle at time $t$ in the actual schedule,
\begin{description}
\item[Step 1.] Use the $\alpha$-queue to compute the next dispatching time $\nextdisp({\cal P}_r, t)$ on processor ${\cal P}_r$ and proceed to the steps 2--5 for every waiting job $\tau_{i,j}$ at time $t$ in the actual schedule.
\item[Step 2.] Compute the amount $L_{i,j}(t)$ of additional time units that $\tau_{i,j}$ could reclaim in the actual schedule if it was dispatched at time $t$, i.e.,
\[ L_{i,j}(t) \equals \min(\nextdisp({\cal P}_r, t), \disp_{i,j}(t)) - t \]
In Figure~\ref{fig:first_rule} we have $\nextdisp({\cal P}_2, 2) = 6$ and $L_{i,j}(2)=4$ $\forall \tau_{i,j}$. 
\item[Step 3.] Compute what would be the resulting execution speed $s_{i,j}'$ if $\tau_{i,j}$ was granted to use both its earliness and these $L_{i,j}$ additional time units, i.e., $s_{i,j}'$ is computed so that $\frac{\rem_{i,j}(t)}{s_{i,j}'} = \frac{\rem_{i,j}(t) + \epsilon_{i,j}(t)}{s_{i,j}^{\offline}} + L_{i,j}$, thus leading to
\[ s_{i,j}' \leftarrow \widehat{S}\left(\frac{\rem_{i,j}(t) \cdot s_{i,j}^{\offline}}{\arem_{i,j}(t) + L_{i,j}(t) \cdot s_{i,j}^{\offline}}\right) \]
\item[Step 4.] Estimate what would be the resulting execution speed $s_{i,j}''$ if $\tau_{i,j}$ was not granted to use these $L_{i,j}(t)$ additional time units. According to Rule~\ref{rule:1}, $s_{i,j}$ will be modified to $s_{i,j}''$ when $\tau_{i,j}$ will be dispatched in the offline schedule (say at time $t''$). By assuming that $\tau_{i,j}$ will not be executed in the actual schedule until time $t''$, we will have $\rem_{i,j}(t'') = \rem_{i,j}(t)$ and from Expression~\ref{equ:speedreduction2}
\[ s_{i,j}'' \leftarrow \widehat{S}\left(\frac{\rem_{i,j}(t) \cdot s_{i,j}^{\offline}}{\arem_{i,j}(t)} \right) \]
\item[Step 5.] Compute the energy saving $\Delta E_{i,j}$ between execution at speed $s_{i,j}''$ and at speed $s_{i,j}'$:
\[ \Delta E_{i,j} \leftarrow E_i\left(\frac{\rem_{i,j}(t)}{s_{i,j}''}, s_{i,j}''\right) - E_i\left(\frac{\rem_{i,j}(t)}{s_{i,j}'}, s_{i,j}'\right) \]
\item[Step 6.] Dispatch the job $\tau_{k,\ell}$ with the largest $\Delta E_{k,\ell}$ to processor ${\cal P}_r$. If $\Delta E_{i,j} \leq 0$ for all the waiting jobs, then dispatch the waiting job $\tau_{k,\ell}$ (if any) with the highest priority in order to complete it earlier and to potentially increase the length of future slack time. 
\item[Step 7.]  If there is a selected job $\tau_{k,\ell}$, set its execution speed $s_{k,\ell}$ to the computed one $s_{k,\ell}'$. Otherwise, turn the processor ${\cal P}_r$ into the idle mode.
\end{description}
\end{Rule}
\end{small}

Notice that, if a processor is about to be idle in the actual schedule exactly when a job is dispatched in the offline one, \emph{only Rule~\ref{rule:1} is applied}. Algorithm~\ref{algo:mora} presents the pseudo-code of $\mora$ and we demonstrate its correctness in the following section. 

\incmargin{.2em} 
\linesnotnumbered
\begin{algorithm}[h!]
\begin{scriptsize}
\lnl{} Determine the offline speed $s_{i,j}^{\offline}$ of every job $\tau_{i,j}$ \;
\lnl{} $\alpha$-queue $\leftarrow \phi$ \;
\vspace{0.2cm}

\textbf{At job arrival (say $\tau_{i,j}$) at time $t$:} 

\Indp
\lnl{} Update the $\alpha$-queue according to $\alpha$-Rule~\ref{arule:4} \;
\lnl{} Insert the value of $C_i$ into the $\alpha$-queue according to $\alpha$-Rule~\ref{arule:3} \;
\lnl{} Set $s_{i,j}$ to $s_{i,j}^{\offline}$ \;
\Indm
\vspace{0.2cm}

\textbf{Whenever any processor ${\cal P}_r$ is about to get idle at time $t$:} 

\Indp
\lnl{} Update the $\alpha$-queue according to $\alpha$-Rule~\ref{arule:4} \;
\lnl{} apply Rule~\ref{rule:2} \;
\Indm
\vspace{0.2cm}

\textbf{Whenever any job $\tau_{i,j}$ is dispatched to any processor ${\cal P}_r$ in the offline schedule at time $t$:} 

\Indp
\lnl{} Update the $\alpha$-queue according to $\alpha$-Rule~\ref{arule:4} \;
\lnl{} \lIf{(a job $\tau_{k,\ell} \neq \tau_{i,j}$ is running on ${\cal P}_r$)}{Preempt $\tau_{k,\ell}$} \;
\lnl{} apply Rule~\ref{rule:1} \;
\Indm
\caption{$\mora$}
\label{algo:mora}
\end{scriptsize}
\end{algorithm}
\decmargin{.2em}

\SubSection{Correctness of $\mora$}
\label{sec:proofs}

In this section, we formally prove that using $\mora$ does not jeopardize the system schedulability.

\begin{Lemma}
\label{lem:wait_run}
Let $S$ be any preemptive and FJP global scheduling algorithm and let $\tau$ be any set of real-time tasks. Suppose that $\tau$ is scheduled by $S$ while using $\mora$, and at time $t$ during the system execution we have $\forall \tau_{i,j}$ and $\forall 0 \leq t' \leq t$:
\[ \rem_{i,j}(t') \leq \arem_{i,j}(t') \]
Then, $\nexists t'$ with $0 \leq t' \leq t$ such that $\exists \tau_{i,j}$ running at time $t'$ in the offline schedule and waiting at time $t'$ in the actual one.
\begin{proof} 
The proof is obtained by contradiction. Suppose that at any time $t'$ such that $0 \leq t' \leq t$, $\exists \tau_{i,j}$ running in the offline schedule and waiting in the actual one. It implies that at time $t'$ in the offline schedule, there are at most $(m-1)$ jobs with an higher priority than $\tau_{i,j}$, whereas there are at least $m$ such jobs in the actual one. In other words, there is at least one job (say $\tau_{k,\ell}$) at time $t'$ with an higher priority than $\tau_{i,j}$, such that $\tau_{k,\ell}$ is completed in the offline schedule, and not in the actual one. For this job, it holds that $\rem_{k,\ell}(t') > \arem_{k,\ell}(t')$, leading to contradiction with our hypothesis. The property follows. \hfill \qed
\end{proof}
\end{Lemma}

\begin{Lemma}
\label{cor:rule1}
Let $S$ be any preemptive and FJP global scheduling algorithm and let $\tau$ be any set of real-time tasks. Suppose that $\tau$ is scheduled by $S$ while using $\mora$, and at time $t$ during the system execution we have $\forall \tau_{i,j}$ and $\forall 0 \leq t' \leq t$:
\[ \rem_{i,j}(t') \leq \arem_{i,j}(t') \]
Then, $\nexists t'$ with $0 \leq t' \leq t$ such that $\exists \tau_{i,j}$ running at time $t'$ in the offline schedule and such that the last speed modification of $\tau_{i,j}$ was performed according to Rule~\ref{rule:2}.
\begin{proof} 
The proof is obtained by contradiction. Suppose that at time $t'$ with $0 \leq t' \leq t$, $\exists \tau_{i,j}$ running at time $t'$ in the offline schedule and such that the last modification of $s_{i,j}$ was performed according to Rule~\ref{rule:2}. Let $t_{\act}$ and $t_{\offline}$ be the largest instants before time $t'$ at which $\tau_{i,j}$ was dispatched in the actual and offline schedule, respectively. Notice that the case where $\tau_{i,j}$ is not dispatched before time $t'$ in the actual schedule leads to a contradiction of Lemma~\ref{lem:wait_run}. Therefore, only two cases may arise: (i) $t_{\act} \leq t_{\offline}$, in this case $s_{i,j}$ would have been modified at time $t_{\offline}$ according to Rule~\ref{rule:1}, leading to a contradiction of our hypothesis, or (ii) $t_{\act} > t_{\offline}$, which leads to a contradiction of Lemma~\ref{lem:wait_run}. The property follows. \hfill\qed
\end{proof}
\end{Lemma}

\begin{Theorem}
Let $S$ be any preemptive and FJP global scheduling algorithm and let $\tau$ be any set of real-time tasks which is schedulable by $S$ when every job $\tau_{i,j}$ is executed at its offline speed $s_{i,j}^{\offline}$. Then, every job deadline is still met when the system is scheduled by $S$ while using $\mora$.
\begin{proof}
The proof consists in showing that $\forall \tau_{i,j}$ we have
\begin{equation}
\label{equ:goal}
\rem_{i,j}(d_{i,j}) \leq \arem_{i,j}(d_{i,j})
\end{equation}
while using $\mora$. Indeed, since the offline schedule meets all the deadlines, we have $\arem_{i,j}(d_{i,j}) = 0$ $\forall \tau_{i,j}$. Therefore, having $\rem_{i,j}(d_{i,j}) \leq \arem_{i,j}(d_{i,j})$ leads to $\rem_{i,j}(d_{i,j}) = 0$ $\forall \tau_{i,j}$, meaning that the actual schedule also meets all the deadlines. 

Initially at time $t=0$, we obviously have $\rem_{i,j}(0) = \arem_{i,j}(0)$ $\forall \tau_{i,j}$. Now, let $t > 0$ be any instant and suppose that $\forall \tau_{i,j}$ and $\forall 0 \leq t' \leq t$ we have $\rem_{i,j}(t') \leq \arem_{i,j}(t')$. We prove in the following that it yields
\begin{equation}
\label{equ:rem_next}
\rem_{i,j}(\next(t)) \leq \arem_{i,j}(\next(t)) \ \ \forall \tau_{i,j}
\end{equation}
where $\next(t)$ denotes the earliest instant after time $t$ such that one of the following events occurs: arrival of a job, deadline of a job, completion of a job in the actual schedule or in the offline schedule, dispatching of a job in the actual schedule or in the offline schedule. Obviously if Inequality~\ref{equ:rem_next} holds then Inequality~\ref{equ:goal} also holds since $\next(t)$ can denote every job deadline.

From the definition of $\next(t)$, every processor of both schedules is either idle or it executes one and only one job during any time interval $\left[ t, \next(t) \right]$. In other words, the state (waiting or running) of any active jobs in any schedule does not change during any time interval $\left[ t, \next(t) \right]$. As a result, the following relations hold at time $t$:
\begin{itemize}
\item For any waiting job $\tau_{i,j}$ in the actual schedule:
\begin{equation}
\label{equ:rel1}
\rem_{i,j}(\next(t)) = \rem_{i,j}(t)
\end{equation}
\item For any waiting job $\tau_{i,j}$ in the offline schedule:
\begin{equation}
\label{equ:rel3}
\arem_{i,j}(\next(t)) = \arem_{i,j}(t)
\end{equation}
\item For any running job $\tau_{i,j}$ in the actual schedule:
\begin{equation}
\label{equ:rel2}
\rem_{i,j}(\next(t)) \leq \rem_{i,j}(t)
\end{equation}
\end{itemize}

The first part of the proof shows that Inequality~\ref{equ:rem_next} holds for every waiting job at time $t$ in the actual schedule and the second part shows that it also holds for every running job at time $t$ in the actual schedule.

\paragraph{Part 1.} Let $\tau_{k,\ell}$ be any waiting job at time $t$ in the actual schedule. From Lemma~\ref{lem:wait_run}, we know that $\tau_{k,\ell}$ is also waiting at time $t$ in the offline one and since by hypothesis $\rem_{k,\ell}(t) \leq \arem_{k,\ell}(t)$, we know from Equalities~\ref{equ:rel1} and~\ref{equ:rel3} that $\rem_{k,\ell}(\next(t)) \leq \arem_{k,\ell}(\next(t))$. The property follows.

\paragraph{Part 2.} Let $\tau_{k,\ell}$ be any running job at time $t$ in the actual schedule. Regarding its execution speed $s_{k,\ell}$, only two cases may occur: its last modification was performed by Rule~\ref{rule:1} (case 1) or by Rule~\ref{rule:2} (case 2).

\paragraph{Case 1.} $\tau_{k,\ell}$ is running at time $t$ in the actual schedule and the last modification of $s_{k,\ell}$ was performed according to Rule~\ref{rule:1} when it was dispatched in the offline schedule (say at time $t_{\offline} \leq t$). By hypothesis, we have $\rem_{k,\ell}(t_{\offline}) \leq \arem_{k,\ell}(t_{\offline})$ and we know that $\tau_{k,\ell}$ is executed non-preemptively in both schedules during the time interval $\left[ t_{\offline}, t\right]$. Indeed, if it was preempted in the offline schedule, it would have been also preempted in the actual one according to Rule~\ref{rule:1}. However in the actual schedule, $\tau_{k,\ell}$ is running at times $t_{\offline}$ and $t$. Therefore, its speed would have been modified according to Rule~\ref{rule:2} at its re-dispatching time if it was preempted during $\left[ t_{\offline}, t\right]$. As a result, from our interpretation of the processor speed we get
\begin{small}
\begin{equation}
\label{equ:temp1}
\rem_{k,\ell}(\next(t)) = \rem_{k,\ell}(t_{\offline}) - s_{k,\ell} \cdot (\next(t) - t_{\offline})
\end{equation}
\end{small}
and
\begin{small}
\begin{equation}
\label{equ:temp2}
\arem_{k,\ell}(\next(t)) = \arem_{k,\ell}(t_{\offline}) - s_{k,\ell}^{\offline} \cdot (\next(t) - t_{\offline})
\end{equation}
\end{small}

\noindent After the speed modification by Rule~\ref{rule:1} at time $t_{\offline}$, we know from Expression~\ref{equ:speedreduction2} that $s_{k,\ell} = \frac{\rem_{k,\ell}(t_{\offline})}{\arem_{k,\ell}(t_{\offline})} \cdot s_{k,\ell}^{\offline}$ and Equality~\ref{equ:temp1} can be rewritten as
\begin{scriptsize}
\[ \rem_{k,\ell}(\next(t)) = \rem_{k,\ell}(t_{\offline}) - \frac{\rem_{k,\ell}(t_{\offline})}{\arem_{k,\ell}(t_{\offline})} \cdot s_{k,\ell}^{\offline} \cdot (\next(t) - t_{\offline}) \]
\end{scriptsize}

\noindent Finally, notice that multiplying the right-hand side of the above Equality by $\frac{\arem_{k,\ell}(t_{\offline})}{\rem_{k,\ell}(t_{\offline})}$ leads to the right-hand side of Equality~\ref{equ:temp2}. Since by hypothesis $\rem_{k,\ell}(t_{\offline}) \leq \arem_{k,\ell}(t_{\offline})$, we have $\frac{\arem_{k,\ell}(t_{\offline})}{\rem_{k,\ell}(t_{\offline})} \geq 1$ and therefore $\rem_{k,\ell}(\next(t)) \leq \arem_{k,\ell}(\next(t))$. The property follows.

\paragraph{Case 2.} $\tau_{k,\ell}$ is running at time $t$ in the actual schedule and the last modification of $s_{k,\ell}$ was performed according to Rule~\ref{rule:2}. Therefore, we know from Lemma~\ref{cor:rule1} that $\tau_{k,\ell}$ is waiting at time $t$ in the offline schedule, and since by hypothesis $\rem_{k,\ell}(t) \leq \arem_{k,\ell}(t)$, we know from Equalities~\ref{equ:rel3} and~\ref{equ:rel2} that $\rem_{k,\ell}(\next(t)) \leq \arem_{k,\ell}(\next(t))$. The theorem follows. \hfill\qed
\end{proof}
\end{Theorem}

\Section{Simulation results}
\label{sec:experiments}

In this section, we compare the effectiveness of $\mora$ with other energy-aware algorithms. However, it is meaningful to only compare $\mora$ with approaches that consider the same models of computation and the most related paper to ours is~\cite{Nelis:08}, where two methods with the same task and platform model are proposed. However, these two methods do not take into account the application-specific parameter $e_i$ of task $\tau_i$. The first method proposed in~\cite{Nelis:08} (that we denote by $\off$ hereafter) is an offline speed determination technique for Global-$\edf$ which determines an unique and constant speed $s^{\offline}$ for all the processors such that all the job deadlines are met under this speed. In our simulations, this $\off$ method is used by $\mora$ in order to provide the offline speed $s_{i,j}^{\offline}$ of every job $\tau_{i,j}$, i.e., $s^{\offline}$ is determined at line 1 of Algorithm~\ref{algo:mora} and $s_{i,j}^{\offline}$ is set to $s^{\offline}$ between lines 4 and 5. The second method proposed in~\cite{Nelis:08} is the $\mote$ algorithm. At run-time, it anticipates the coming idle instants in the schedule and adjusts the speed of the processors accordingly, i.e., it reduces the processors speed in order to minimize the proportion of time during which the system is idle. Since this algorithm is also based on the concept of the offline speeds, we consider that $\off$ is also used to provide it. Although $\mora$ could also be compared with frame-based scheduling algorithms (since the sporadic task model is a generalization of the frame-based task model), we do not perform such comparisons in this paper.

In our simulations, we schedule \emph{periodic implicit-deadline systems} (i.e., $\forall \tau_i$, $T_i$ is here the \emph{exact} inter-arrival delay between successive jobs and $D_i = T_i$). The energy consumption of each generated system is computed by simulating three methods: $\mote$, $\mora$ and $\moraote$, i.e., a combination of the $\mote$ and $\mora$. Indeed, since these algorithms do not interfere with each other, the $\mote$ rule can be applied \emph{on the offline speeds} just \emph{before applying Rule~\ref{rule:1}} of $\mora$ (i.e., between lines 9 and 10 of Algorithm~\ref{algo:mora}). Although the implementation details of $\moraote$ are omitted here due to the space limitation, we will see in our simulation results that this combination always improves the provided energy savings. The consumptions provided by these three methods are compared with the consumption of the $\maxmethod$ method, where all the jobs are executed at the maximal processors speed $s_{\max} = 1$. That is, we consider that the consumption by $\maxmethod$ is $100\%$ and the consumptions of the other methods are normalized.

In every simulation, we generated $100$ set of tasks with a total density $\delta_{\somme}(\tau)$ within $\left[ d, d+0.05\right]$ where $d=0, 0.05, \ldots, 9.95$, leading to an amount of $20000$ generated task sets for each simulation. The upper bound on $\delta_{\somme}(\tau)$ (i.e., 10) was chosen in order to cover a large number of systems while keeping the simulation time reasonable. For a given \emph{total density}, tasks densities $\delta_i$ are uniformly generated within $\left[ 0.01, D_{\max}\right]$ until the total density $\delta_{\somme}(\tau)$ reaches the expected one (the upper bound $D_{\max}$ on the tasks density will be discussed later). Notice that the number $n$ of tasks is not fixed beforehand, i.e., it depends on this step that generates task densities. Next, other task parameters $C_i$, $D_i$ and $T_i$ are randomly generated according to their respective density $\delta_i$. Finally, the application-specific parameters $e_i$ are uniformly chosen in $\left[ 0.8, 1.2 \right]$ so that the consumption of the tasks varies between $80\%$ and $120\%$ of the power of the measured benchmark. 

Once a set of tasks is generated, it is executed during $100$ hyper-periods (i.e. the least common multiple of the task periods) by the four methods $\maxmethod$, $\mote$, $\mora$ and $\moraote$. This upper bound on systems execution time was chosen to ensure that every task generates at least $100$ jobs (for the same reason as those mentioned above). During each system execution, the actual execution time of every job $\tau_{i,j}$ is uniformly generated in $\left[ \frac{C_i}{10}, C_i\right]$. This lower bound $\frac{C_i}{10}$ was chosen in order to reflect the fact that a job may take up to $10$ times less than its WCET. Finally, for every generated task set $\tau$, the number $m$ of processors must be sufficient to schedule $\tau$ by $\maxmethod$ without missing any deadline. Hence, we set $m$ to the lowest integer that passes one of the following $\edf$-schedulability tests: the \emph{density-based} test~\cite{Goossens2003Priority-driven}, the \emph{load-based} test~\cite{BaruahBaker:08} and the test denoted Test 13 in~\cite{LivreBaruah}. Simulations were performed while considering different scheduling algorithms (Global-$\edf$ and Global-$\dm$) and various processor models. However, due to the space limitation, we only depict in this paper the results provided by Global-$\edf$ on Intel XScale processors (outlined in Table 1 page~\pageref{tab:processor_characterisitc}). 

\begin{Observation}
\label{obs:2}
The effectiveness of both $\mora$ and $\mote$ mainly relies on the ratio $\frac{m}{n}$, but antagonistically. 
\end{Observation}

This observation stems from the fact that $\mora$ saves energy via the waiting jobs whereas $\mote$ profits from the absence of waiting jobs. When $\frac{m}{n}$ tends to $1$, jobs tend to never wait for a free processor and $\mote$ therefore provides significant energy savings whereas the effectiveness of $\mora$ is almost null. On the other hand when $\frac{m}{n}$ tends to $0$, processors tend to consecutively execute several distinct jobs and jobs are often waiting. As a result, $\mora$ is often able to reclaim unused time and provides important energy savings whereas the effectiveness of $\mote$ is negligible.

According to our task generation process, we are not able to directly set the ratio $\frac{m}{n}$ to any given value. However, the number $m$ of processors is obtained by using a combination of \emph{sufficient} schedulability tests and the accuracy of these tests mainly relies on $\delta_{\max}(\tau)$. Basically, the ratio $\frac{m}{n}$ increases as $\delta_{\max}(\tau)$ becomes larger and since the generated task sets are more likely to have a large $\delta_{\max}(\tau)$ when the upper bound $D_{\max}$ is high, we can indirectly control the ratio $\frac{m}{n}$ via $D_{\max}$. The Y-axis of Figure 3 represents the ratio $\frac{m}{n}$ obtained from the used schedulability tests when $\delta_{\somme}(\tau)$ varies within $\left[ 0, 10 \right]$ and $D_{\max}$ varies within $\left[ 0.1, 1 \right]$ with a step of $0.1$. 

For every $D_{\max}$ multiple of $0.1$ within $\left[ 0.1, 1 \right]$, $20000$ set of tasks were generated by the generation process described above and the resulting average consumptions of the $\mote$, $\mora$ and $\moraote$ are depicted in Figure 4. The Y-axis is the average energy consumption of every method compared with the $\maxmethod$ method (in $\%$) and the X-axis is the corresponding value of $D_{\max}$ during the simulation. As we can see, Figures 3 and 4 clearly corroborate Observation~\ref{obs:2}. Moreover, Figure 4 shows that $\mora$ can save up to $32\%$ of energy (in average) over the $\maxmethod$ method (for $D_{\max} = 0.1$) and the algorithm $\moraote$ provides important energy savings for various values of $D_{\max}$. Notice that a part of the energy savings is explained by the use of the $\off$ method, which leads $\mote$ and $\mora$ to an energy savings of about $10\%$ when $\frac{m}{n}$ tends to 0 and 1, respectively. Furthermore, although other processor models and scheduling algorithms led to different average consumptions, the \emph{evolution} of the consumption with respect to $D_{\max}$ remains similar than in Figure 4.

\Section{Conclusion}
\label{sec:conclusion}

In this paper, we propose a slack reclamation scheme called $\mora$ which reduces the energy consumption while scheduling a set of \emph{sporadic constrained-deadline tasks} by a \emph{global}, \emph{preemptive} and FJP algorithm on a \emph{fixed number of DVFS-identical processors}. According to~\cite{Chen:07} and to the best of our knowledge, we are the firsts to address such approach in this context. The proposed algorithm $\mora$ exploits early job completions at run-time by starting the execution of the next waiting jobs at a lower speed. Compared with other reclaiming algorithms such that the $\dra$ proposed in~\cite{AyMeMoMe:04}, $\mora$ takes into account the application-specific consumption profile of the tasks in order to improve the energy saving that it provides. Moreover, we proved that using $\mora$ does not jeopardize the system schedulability and we show in our simulations that it can save up to $32\%$ of energy (in average) compared to execution without using any energy-aware algorithm. 

\begin{figure}[h!]
\begin{center}
\includegraphics[height=.3\textwidth,width=.5\textwidth]{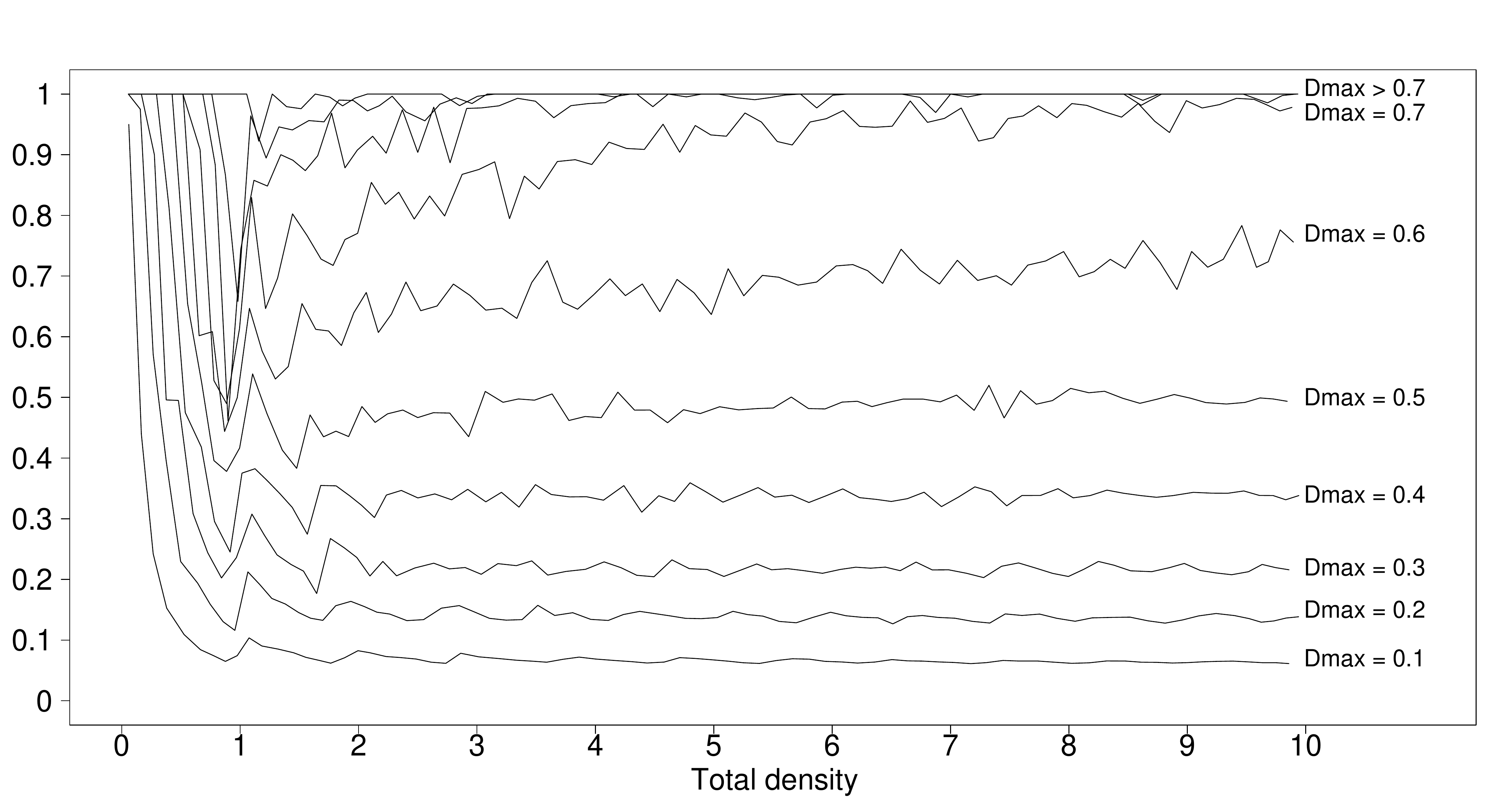}
\label{fig:ratios}
\vspace{-5ex}
\caption{Ratio $\frac{m}{n}$ for different values of $D_{\max}$}
\vspace{-3ex}
\end{center}
\end{figure}

\begin{figure}[h!]
\begin{center}
\includegraphics[height=.3\textwidth,width=.5\textwidth]{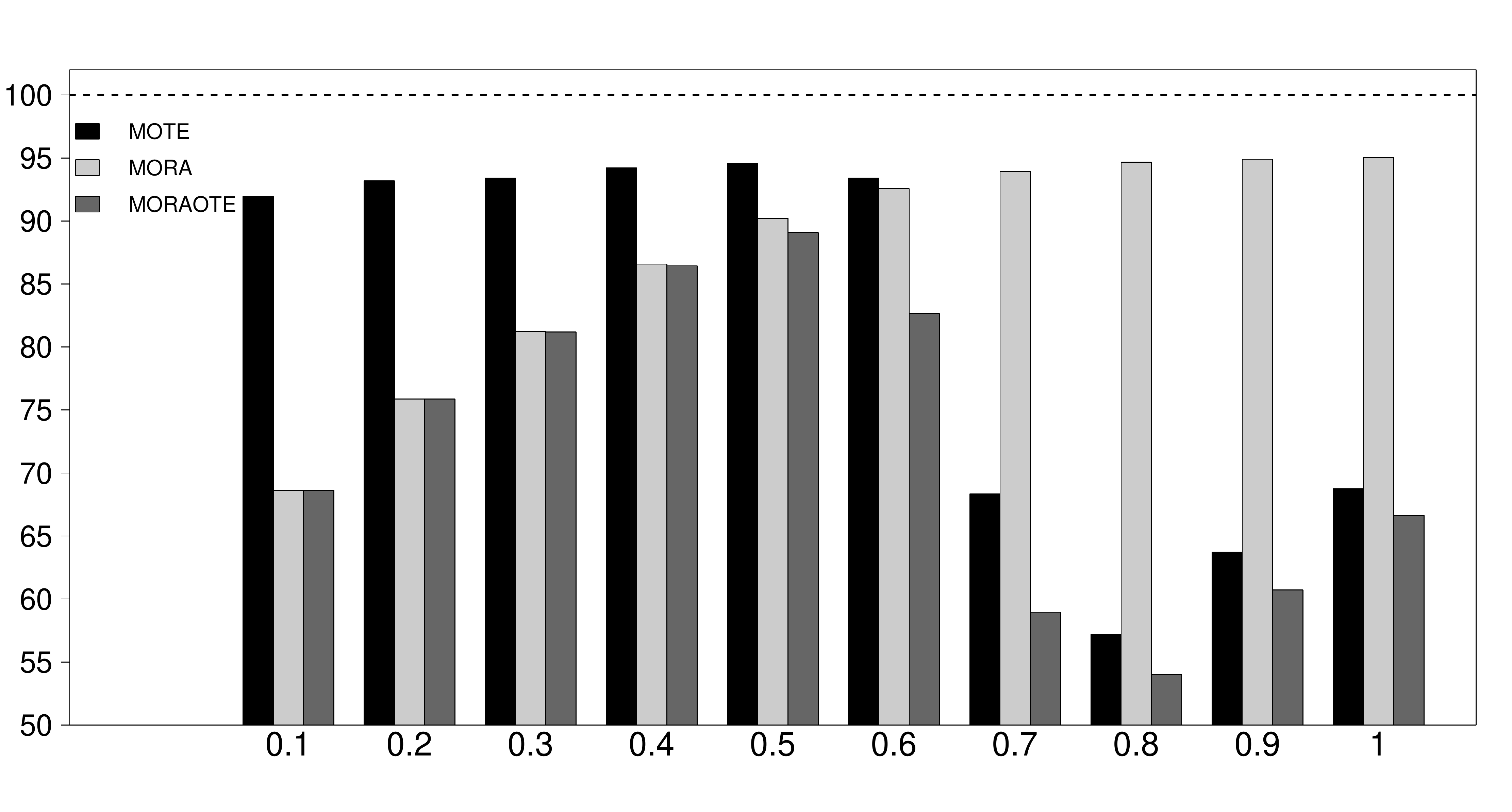}
\label{fig:hist}
\vspace{-5ex}
\caption{Average consumptions of $\mote$, $\mora$ and $\moraote$ for various values of $D_{\max}$ under Global-$\edf$ on Intel XScale processors.}
\end{center}
\end{figure}

In our future works, we aim to specialize $\mora$ so that it will take into account more practical constraints such that preemption costs, migration costs and time overheads due to the multiple frequency switching. Moreover, we aim to extend our processor model in order to handle the various idle and sleep modes of the processors and to take into account the energy costs due to frequency switching. In other future works, we also aim to propose a new multiprocessor reclamation scheme which anticipates the early completion of jobs for further reducing the CPU speed. This approach will be based on statistical informations about tasks that are assumed to be known \emph{a priori}. Some \emph{uni}processor energy-aware algorithms already exploit this concept (see the $\agr$ algorithm proposed in~\cite{AyMeMoMe:04} for instance).

\bibliographystyle{latex8}
\bibliography{energy}

\end{document}